\documentclass[10pt, twocolumn]{article}
\usepackage{latex8}
\usepackage[small,compact]{titlesec}
\usepackage{algorithmic}
\usepackage{algorithm}
\usepackage{times}
\usepackage{latexsym}
\usepackage{setspace}
\usepackage[hmargin={0.95in,0.95in},vmargin={0.95in,0.945in}]{geometry}
\usepackage{graphicx}
\usepackage{subfigure}
\usepackage{balance}

\title{Using Dedicated and Opportunistic Networks in Synergy for a
 Cost-effective Distributed Stream Processing Platform}

\author{
\textit{Shah Asaduzzaman} \\
School of Information Technology and Engineering \\
University of Ottawa\\
Ottawa, ON, K1N 6N5, Canada \\
Email: \texttt{asad@site.uottawa.ca}
\and 
\textit{Muthucumaru Maheswaran} \\
School of Computer Science \\
McGill University \\
Montreal, QC H3A 2A7, Canada \\
Email: \texttt{maheswar@cs.mcgill.ca}
}

\date{}

\begin{document}


\pagestyle{empty}
\maketitle
\thispagestyle{empty}

\begin{abstract}
  This paper presents a case for exploiting the synergy of dedicated
  and opportunistic network resources in a distributed hosting
  platform for data stream processing applications. Our previous
  studies have demonstrated the benefits of combining dedicated
  reliable resources with opportunistic resources in case of
  high-throughput computing applications, where timely allocation of
  the processing units is the primary concern. Since distributed
  stream processing applications demand large volume of data
  transmission between the processing sites at a consistent rate,
  adequate control over the network resources is important here to
  assure a steady flow of processing. In this paper, we propose a
  system model for the hybrid hosting platform where stream processing
  servers installed at distributed sites are interconnected with a
  combination of dedicated links and public Internet. Decentralized
  algorithms have been developed for allocation of the two classes of
  network resources among the competing tasks with an objective
  towards higher task throughput and better utilization of expensive
  dedicated resources.  Results from extensive simulation study show
  that with proper management, systems exploiting the synergy of
  dedicated and opportunistic resources yield considerably higher task
  throughput and thus, higher return on investment over the systems
  solely using expensive dedicated resources.
\end{abstract}

\section{Introduction}
\label{sec:intro}
Many applications on the Internet are creating, manipulating, and
consuming data at an astonishing rate. Data stream processing is one
such class of applications where data is streamed through a network of
servers that operate on the data as they pass through
them~\cite{Pietzuch2006, Seshadri2007, Gates2004}. Depending on the
application, the stream processing tasks can have complex topologies
with multiple sources or multiple sinks. Examples of stream processing
tasks are found in many areas including distributed databases, sensor
networks, and multimedia computing. Some examples include: (i)
multimedia streams of real-time events that are transcoded into
different formats, (ii) insertion of information tickers into
multimedia streams, (iii) real-time analysis of network monitoring
data streams for malicious activity detection, and (iv) function
computation over data feeds obtained from sensor networks.

One of the salient characteristics of this class of applications is
the demanding compute and network resource
requirements~\cite{Kalogeraki2007}. Huge volume of data generated at a
high rate need to be processed within real-time constraints. Moreover,
various operations on these data streams are provided by different
servers at distributed geographic locations~\cite{Nahrstedt2006}. All
these factors demand a scalable and adaptive architecture for
distributed stream processing platform, where fine-grained control
over processing and network resources is possible.

Earlier works on stream processing engines~\cite{STREAM2003}
resorted to centralized single-server or server-cluster based
solutions where tighter control over available resources are
possible. With possibility of different processing services or
operations being provided by different providers, need for distributed
stream processing platform arose. Several architectures have been
proposed to support such distributed processing of
streams~\cite{Kalogeraki2007, Schwan2005, Nahrstedt2006,
Karamcheti2004}. Due to the stringent rate-requirement for processing
and transmission of data, most researchers have assumed a central
resource controller that can gather the availability status of all
resources and map the requested tasks on them. However, with advent of
diverse range of stream processing services, it is important to allow
autonomous providers of services to collaborate and share their
resources. Thus it is important to develop distributed resource
allocation schemes, where control is available over local resources
only.

While it is feasible to have dedicated server resources and precisely
allocate them for processing tasks, dedicated networks over wide-area
installations remain costly. It is possible to propagate the
data streams through the the distributed servers using the public
Internet. However, the lack of adequate control over end-to-end bandwidth in
current Internet and the stringent rate requirement of the stream
processing applications demand some dedicated network resources. In
fact, recent advances in optical network technologies such as
user-controlled light path~\cite{Boutaba2008} open the
possibility of on-demand provisioning of end-to-end optical links with
total control of the bandwidth available to the user application.

In this paper, we explore a novel approach where a combination of
dedicated links and public network to interconnect the servers. The
main focus of this paper is to explore how such a hybrid (denoted {\em
bi-modal} in this paper) network can be best used for data stream
processing tasks. The hypothesis that drives this work is that the
combination has a synergistic effect that allows better utilization of
the dedicated resources, and yields higher return on investment. We
devised distributed algorithms for allocation of these hybrid
resources to demonstrate the viability of this synergy hypothesis.

This paper extends some of our previous work \cite{jpdc07,JPDC06} on
bi-modal compute platforms where dedicated compute-clusters were
augmented with opportunistically harvested processing elements to
increase work throughput and utilization of dedicated resources. Using
data stream processing tasks as a concrete example, this paper
demonstrates the benefit of using bi-modal network infrastructures for
communication-intensive applications. 

In Section \ref{sec:model} we present the system model for the data
stream processing and the associated resource allocation
problem. Section \ref{sec:problem} discusses the algorithms devised
for managing the resources towards global optimization of throughput
and resource utilization. Section \ref{sec:results} examines the
results from the extensive simulation studies we carried out to
evaluate the algorithms. Section \ref{sec:related} reviews related
literature.

\section{System Model and Assumptions}
\label{sec:model}
In a {\em stream processing task}, the data stream originating from a
{\em data-source} node, progresses through several steps of prcessing,
termed as {\em service components} (or {\em service} in short), before
being delivered to the {\em data-delivery} node. For example, in video
streaming, the service components may be encoding of video, embedding
some real time tickers and transcoding the video into different
formats. Although, in very general terms, the data-flow topology could
be arbitrary graphs, in this paper, we restrict our study within
linear path topology only.

The distributed stream procesing platform consists of several
autonomous server nodes that serve the service components. A single
server may serve multiple services and a serve may be available at
multiple servers. Several pairs of servers establish dedicated
point-to-point links between them to have the flow of the data streams
at a controlled rate. Each server is also connected to the public
Internet and end-to-end TCP connection can be established between any
pair of servers. However, end-to-end bandwidth of the TCP connections
cannot be allocated and the flow rate cannot be controlled.

The platform is modeled as an asynchronous message passing distributed
system, where there is no centralized controller to coordinate the
resources. The servers have knowledge of and can precisely allocate
the local resources only, i.e. the processing capacity and the
bandwidth of outgoing links. However, the servers comply with the
global protocol and respond to a predefined set of messages in a
predefined way. The objective of the global protocol is to ensure
adequate resources for each individual task for its seamless progress,
and to maximize the global work throughput. Other factors such as
balancing the load among different servers and maximizing the
utilization of dedicated resources are also considered. Design and
evaluation of the protocol constitute the remaining sections of the
paper.

\begin{figure}[htbp]
    \centering
    \includegraphics{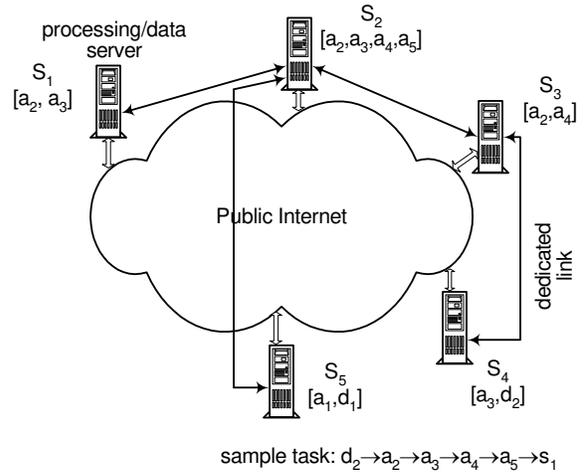}
    \caption{Stream processing platform}
   \label{fig:arch_stream_arch}
\end{figure}

Figure~\ref{fig:arch_stream_arch} illustrates a scenario of a stream
processing platform containing five servers. The example stream
processing task shown in the figure requests a data stream from data
source $d_2$ to be processed through services $a_2$, $a_3$, $a_4$ and
$a_5$, and to be delivered to $S_1$. This task
may be served by the servers $S_4$ (serving $d_2$), $S_3$ (serving
$a_2$), $S_2$ (serving $a_3$ and $a_4$). Either dedicated link or
public network may be used to transmit the data stream between any two
consecutive servers.

For convenience, the resource allocation process is divided into two
phases. First, individual tasks with multiple service components are
mapped on the processing servers fulfilling the processing and
transport capacity requirements. A cost function is used to select the
best among multiple feasible maps. The second phase re-allocates the
link bandwidths among competing tasks, after the tasks start execution
based on the initial allocation. This is necessary because of the
variability of data rate in the end-to-end TCP connections on public
Internet. Both the re-allocation phases and intial allocation are
driven by the same global optimization goal, namely maximization of
global throughput and resource utilization, subject to fulfillment of
individual task requirements.

The specification of the stream processing task includes the ordered
sequence of service components, the data source node, the data
delivery node and the desired rate of data delivery. We assume a rate
based model~\cite{Kalogeraki2007, Karamcheti2004} to specify resource
requirement for each service component. For any service, both the
output data rate and the CPU requiremnt are proportional to the input
data rate, and are specified by two factors -- the {\em bandwidth
shrinkage factor} and the {\em CPU usage factor}, respectively. We
also assume a rate based pricing for the services. The task
specification includes a price per byte of data delivered. This is
directly translated to apportioned prices for each of the service
components, using the above two factors. The task specification is a
{\em service level agreement} (SLA) between the user and the platform.

\section{Decentralized Management of Server and Network Resources}
\label{sec:problem}
A resource manegement engine (denoted as RMS agent) runs in each
server that implements the protocols for coordinated allocation of
network and CPU resources. Each RMS agent has two modules -- a map
manager and a dynamic scheduler, to perform the two phases of recource
allocation described before. This section describes the algorithms
that encodes the functions of these two modules in details.

A user of the distributed platform uses one of the server nodes as a
portal to launch its stream processing task. The portal node then
engages the map manager to initiate the mapping of the specified
requirements on the network. Through message passing among the map
managers in different server nodes, the distributed mapping algorithm
results in a set of feasible maps at the map manager of the
data-source node. Each of the maps defines a path from the data source
node to the delivery node through the server nodes that serve
necessary service components. The best among the available feasible
maps according to a certain cost metric is selected.

\begin{figure*}[htb]
\begin{minipage}[b]{0.48\linewidth}
\centering \includegraphics{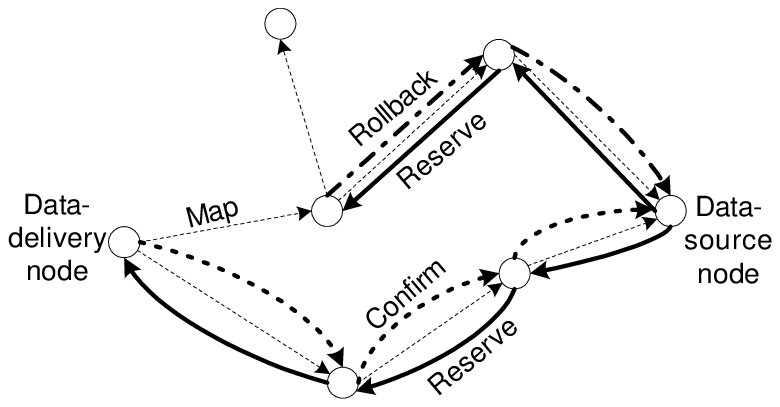}
\caption{Mapping, reservation and rollback}
\label{fig:map_reserve}
\end{minipage}
\hspace{0.5cm}
\begin{minipage}[b]{0.48\linewidth}
\centering
\includegraphics{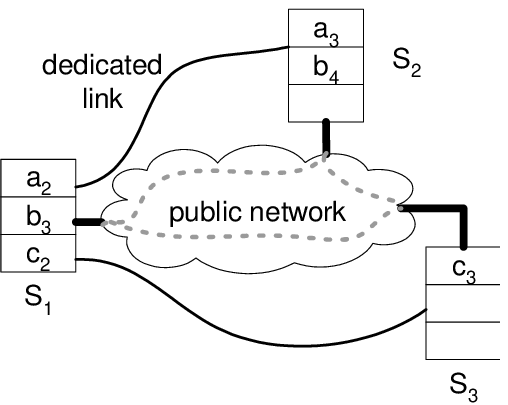}
\caption{Dynamic scheduling of link resources done by server $S_1$ on
  three competing task segments $a_2$-$a_3$, $b_3$-$b_4$, $c_2$-$c_3$}
\label{fig:dynasched}
\end{minipage}
\end{figure*}

A reservation probe is sent from the data-source node to the
data-delivery node along the path found in the selected map. The RMS
agent at each server node along the path tries to allocate the server
and link resources prescribed by the map. Because the mapping process
for multiple tasks may be ongoing concurrently, it is possible that
the required resource is no longer available. In such case the
allocation fails, the probe is rolled back and the next feasible map
is probed by the data-source node. Once a successful probe reaches the
data-delivery node at the other end, a confirmation is sent back to
the data-source node to begin the streaming. The message flow of
mapping and reservation is illustrated in
Figure~\ref{fig:map_reserve}.

The dynamic scheduler module in each server node periodically
re-allocates the locally available link resources among the competing
tasks that are using that server node. The re-allocation process is
illustrated in Figure~\ref{fig:dynasched}. The re-allocation is done
with two objectives -- improving the compliance to the SLA defined
data-delivery rate and maximizing the global processing
throughput. Note that only link resources are re-allocated while
keeping the allocation of server resources unmodified. This follows
from the assumption that servers are dedicated and their processing
rates do not vary over time.

\subsection{Distributed Algorithm for Mapping}
\label{subsubsec:problem_mapreserve_algo}
The problem of mapping a stream processing task specification on
arbitrary network of server nodes subject to processing capacity and
bandwidth constraints is an NP-Complete
problem~\cite{HPCS2007}. Detailed analysis of the problem and
algorithms to solve it in both centralized and distributed manner can
be found in~\cite{AsadPhDThesis, HPCS2007}. Algorithm for a similar
mapping problem was also discussed in~\cite{Nahrstedt2006}. However,
the problems discussed in the above references do not consider
bi-modal network links. Here, we adopt the distributed mapping
algorithm presented in~\cite{HPCS2007} with modifications to
accommodate bi-modal network links.

The distributed mapping of the task specification performed by
gradually expanding the maps to neighbors in the server network.  The
portal server initiates the algorithm by generating the initial map
message.  The {\em ProcessMap} algorithm described in
Algorithm~\ref{alg:pathmap_distr} is executed by the map manager at
each server node on receiving a map message.

\begin{algorithm}[h]
    \begin{algorithmic}[1]
       \STATE {\label{line_input}{\bf Input:} Map message $m$ containing the mapping of first 
	 $j$ services on a series of server nodes is received by node
	 $u$. $j$ is called the {\em prefix-length} of $m$. $T$
	 denotes the ordered set of services in the task}
       \IF{\label{line_2}$u$ is the data-source node and all the services except the
          data source is mapped in $m$}
       \STATE{$m$ is a feasible map}
       \ELSE
          \FOR{$x=0$ to $|T|-j-1$} 
             \IF {service $j+x$ is provided by node $u$ and node
               capacity permits the required processing rate}
	        \STATE{$m_x = $ map found by extending next $x$ services
                  in $T$ on $u$}
             \ELSE
                \STATE{break}
             \ENDIF
	     \FOR {each node $v$ such that there is a dedicated link $(u,v)$}
	        \IF {(available bandwidth in $(u,v)$ link $\geq$ the bandwidth need for the service hop $(j+x, j+x+1)$)
                and budget allows the extension of $m_x$ to $v$} 
	        \STATE{Send $m_x$ to $v$}
		\ENDIF
	     \ENDFOR 
             \IF {$x > 0$} 
                \FOR {each node $v$ such that $v$ provide the service $j+x+1$} 
                   \IF {available uplink bandwidth to the Internet
                     $\geq$ bandwidth need for service hop $(j+x,
                     j+x+1)$} 
                        \STATE{Send $m_x$ to $v$}
                    \ENDIF
                \ENDFOR
             \ENDIF
          \ENDFOR
       \ENDIF
   \end{algorithmic}
    \caption{\em ProcessMap(u, m, T)}
    \label{alg:pathmap_distr}
\end{algorithm}

The algorithm first extends the received partial map by mapping next
few service components on itself as long as the service is available
and processing capacity permits (line 6--8). Each possible extension
is then sent to neighboring nodes subject to availability of network
bandwidth (line 11--22). Note that it is possible to extend the map to
the neighbors without having any service component mapped on the
current server. This allows multi-hop connection between nodes
processing consecutive services. This is beneficial in cases where
there is no direct dedicated link between two server nodes. 

In case of links through the public network, such multi-hops are
unnecessary, because overlay link can be established between any pair
of nodes. This case is handled in lines 16--22. Note that end-to-end
bandwidth cannot be allocated in case of public network links, only
uplink bandwidth can be controlled. In case of extending through a
public network link, only the nodes that provide the service required
in the next hop is chosen (line 17). We assume that an underlying
gossip like algorithm disseminates the presence of services in each
server across the network. Thus, for each service type, each node has
the knowledge of (possibly a subset of) the nodes that hosts that
service. Incorrectness of this information does not cause any
inconsistent mapping, only some feasible maps are missed.

Cyclic mapping is allowed in the extension in lines 12--14. Because
$x=0$ is allowed, it is possible that the map grows to an infinite
length. In practice, this is avoided by limiting the growth of the
multi-hop mapping using a budget factor. Based on the
price-per-byte-processed quoted in the SLA, the allocated revenue for
processing of the $j-th$ service is limited. When the output of the
$j$-th service is sent to the server providing $(j+1)$th service using
a dedicated link, host of the $j$-th service needs to pay and thus
loses revenue. The cost of transmission grows as more dedicated links
are used in a multi-hop link to send the same data. Thus the number of
hops in such multi-hop links are limited by the revenue budgeted for
the service and cost of each hop of dedicated connection.

Because the algorithm enumerates all feasible maps, it generates an
exponential number of messages. Some simple heuristics can limit the
complexity without sacrificing much of optimality. We have used a
simple heuristic called {\em LeastCostMap} where each node remembers
the lowest cost map it has observed so far for each possible prefix
length (number of components already mapped), and it does not extend a
map with higher cost for the same prefix length. Evaluation of
performance of the heuristic compared to other possible heuristics can
be found in~\cite{AsadPhDThesis}.

To devise an appropriate cost metric for choosing the best mapping
among alternative feasible maps, we considered the following two
factors - balancing the service workload among the servers and
minimizing the uncertainty of using public network links where a
dedicated link is available. The load-balance factor for a map (or a
partial map) is computed as an average of the server load-factors
(ratio of used capacity to total capacity) for all the servers
included in the map, and is always a number between $0$ and $1$. A map
with lower load-balance factor spreads the components of a task on
different servers rather than putting all of them into one, and
chooses the under-utilized servers.  In case two maps have almost same
load-balance factor, (do not differ by more than $0.1$ or $10\%$),
then the one in which the number of hops (links connecting the
processing components) assigned to dedicated links is higher is
preferred. If that is also same, the map in which less number of hops
are assigned to public network link is preferred.

\subsection{Algorithm for Dynamic Link Re-allocation}
\label{subsec:problem_dynasched}
The dynamic link scheduler in each server node is invoked periodically
at regular intervals. Based on current evaluation of locally observed
performance, the scheduler re-allocates the locally available link
resources among the competing tasks that are using this server
node. The overall policy of the scheduler is to prioritize the tasks
for use of the network links, based on their deviation from target
data rate and the price they would pay for the data processing
service.

The links that carry the stream between two data processing servers
can be of three different types -- i) a direct dedicated link, ii) a
multi-hop dedicated link through one or more forwarding nodes iii) an
overlay link through the public network. A mapping of a task may
contain any combination of these three types of links between the
processing nodes. Among them, the direct dedicated links are the most
preferred one, because they provide controlled and stable data rate. A
multi-hop dedicated link provides similar control and stability, but
it costs more (Section~\ref{subsubsec:problem_mapreserve_algo}).  The
third possibility is having an overlay link through the public
network. The flow rate is variable over such links, but there is no
additional per-byte cost for sending data through them. So, the nodes
try to opportunistically use these links when dedicated links are
overloaded or not available.

\begin{algorithm}[htbp]
    \begin{algorithmic}[1]
      \STATE{Invoked for each node $u$ periodically}
      \STATE{Group the tasks that are being processed in $u$ by their
      next hop server $v$}
      \FOR{Each group $v$}
           \STATE{Compute the priority of each flow competing for a
      ($u$,$v$) link as -}
	   \STATE{priority $\leftarrow$ budget per byte of processed data *
      bandwidth required to comply with the target rate}
	   \IF {any dedicated link ($u$,$v$) exists}
	   \STATE{Assign the dedicated link to top priority flows
      until all capacity is used}
	   \ENDIF
	   \STATE{Collect all the unassigned flows} 
      \ENDFOR
      \FOR{ All the remaining flows}
       \IF{The budget permits $k$-hop ($u$,$v$) dedicated link, $k>1$}
       \STATE{Launch a probe search and reserve multi-hop dedicated
      path for the flow with maximum $k$ hops}
       \STATE{Assign public network bandwidth for the flow temporarily}
       \ELSE
       \STATE{Assign public network bandwidth for the flow}
       \ENDIF
      \ENDFOR
   \end{algorithmic}  
   \caption{Link re-allocation algorithm}
   \label{alg:netdyna_schedalgo}
\end{algorithm}

Algorithm~\ref{alg:netdyna_schedalgo} is executed when the scheduler
is inviked at reguler intervals. For allocation of the links, tasks
are grouped according to their next hop server node (Line 2). While
prioritizing among competing tasks for each group (Lines 4-5), the
scheduler tries to maximize the revenue earning of the server and
prefers the tasks marked with higher price per unit of processing.  On
the other hand, the servers get penalized on the revenue, if they do
not deliver the processed stream at the agreed upon rate. Therefore
each server tries to fulfill the rate requirements of each task as
much as possible. Thus, the task that requires more bandwidth to
comply with its target gets higher preference. Hence the scheduler
computes the priority of each task as a product of the apportioned
price and the data rate required in next scheduling epoch.

For each next hop group, highest priority tasks get allocation from
the direct dedicated link, if such link exist and capacity permits
(Lines 6-8). The next prior tasks are assigned multi-hop dedicated
links (Lines 12-14). The maximum possible hops in such multi-hop links
are restricted by the apportioned price for that service according to
the task specification. The flows of the remaining tasks from all the
groups are allocated bandwidth from the public overlay links (Lines
15-17).

\section{Performance Evaluation and Discussion}
\label{sec:results}
\subsection{Simulation Model}
\label{subsec:netdyna_sim_model}
We constructed a simulation model of the proposed distributed stream
processing platform using Java based discrete event simulator
JiST~\cite{Jist2005}. Each server in the platform is connected to the
Internet using last mile bandwidth between $1$ Mbps and $2$ Mbps,
randomly assigned. To model the variability of data rate on end-to-end
Internet paths, we used the statistics presented by Wallerich and
Feldmann~\cite{Wallerich2006}. From their data collected from packet
level traces from core routers of two major ISPs over 24 hours, the
logarithm of the ratio of the observed transient flow rate to the mean
flow rate over long period is almost a Normal distribution.
In our simulations, all flows on the public network are perturbed
every $10$ milliseconds according to this model. With the allocated
bandwidth as the mean rate and the standard deviation of the log-ratio
set at $1$, in $95\%$ of the cases the observed bandwidth remains
between one fourth ($2^{-2\sigma}$) and four time ($2^{2\sigma}$) of
the allocated or mean bandwidth.

In addition to the public network links, the servers are
interconnected through dedicated links (which may be leased lines or
privately installed links). For the dedicated network, we assume a
preferential connectivity based network growth model similar to the one
proposed by Barabasi et al~\cite{Barabasi1999}. The basic premise here
is that when a server attempts to establish a dedicated link, it does
so preferably with the most connected server. This eventually results
in a power law degree distribution in the network. We assumed that
server CPU capacity is proportional to the number of dedicated links
it has. The variety of services that a server can host is also
proportional to the node degree or capacity.  The dedicated links have
much higher bandwidth than the network links connecting a node to the
public network. Their bandwidths were randomly assigned between $1$
Mbps and $10$ Mbps and the propagation delays were assumed to be
between $1$ and $10$ milliseconds.  The propagation delay of an
end-to-end connection through the public network was much higher and
assumed to be between $10$ and $100$ milliseconds.

\begin{figure*}[htb]
  \centering
  \setcounter{subfigure}{0}
  \subfigure[Task acceptance ratio]{
    \label{fig:netdyna_acceptratio_numlinks}
    \includegraphics[scale=0.98]{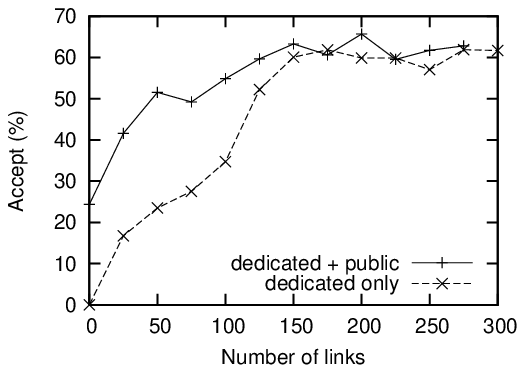}
  }
  \subfigure[Server utilization at different workload]{
    \label{fig:netdyna_cpuutilnet_arvrate}
    \includegraphics[scale=0.98]{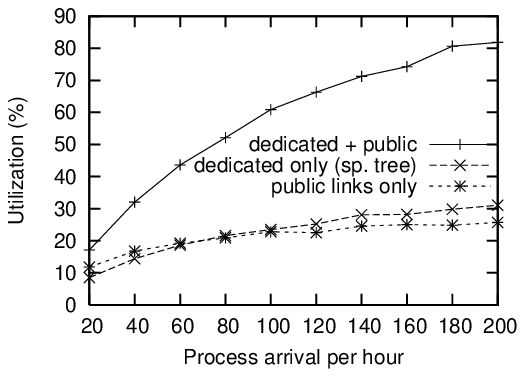}
  }
  \subfigure[Link utilization at different workload]{
    \label{fig:netdyna_dedutilnet_arvrate}
    \includegraphics[scale=0.98]{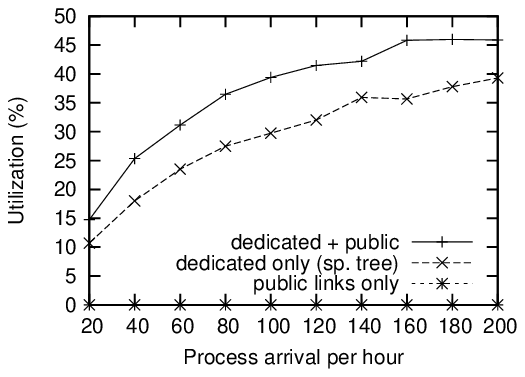}
  }
  \vspace{.1cm}
  \subfigure[Link utilization vs number of dedicated links]{
    \label{fig:netdyna_dedutilnet_numlinks}
    \includegraphics[scale=0.98]{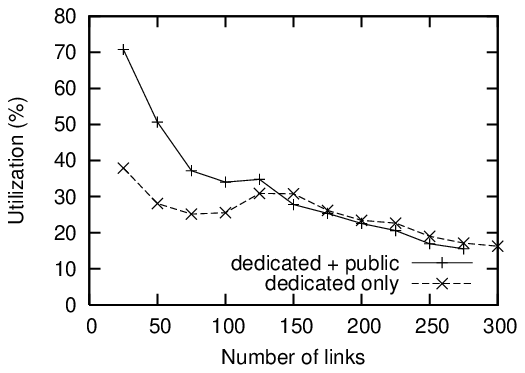}
  }
  \subfigure[SLA deviation vs number of dedicated links]{
    \label{fig:netdyna_deviation_numlinks}
    \includegraphics[scale=0.98]{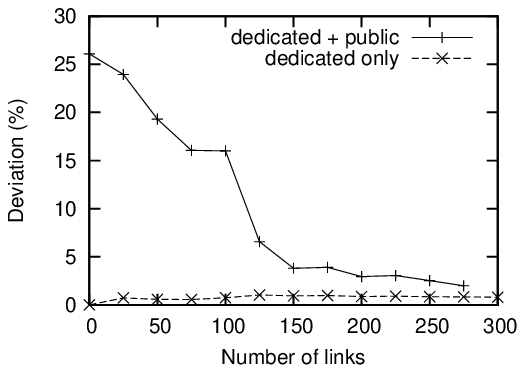}
  }
  \subfigure[Elongation of task execution time]{
    \label{fig:netdyna_exectime_arvrate}
    \includegraphics[scale=0.98]{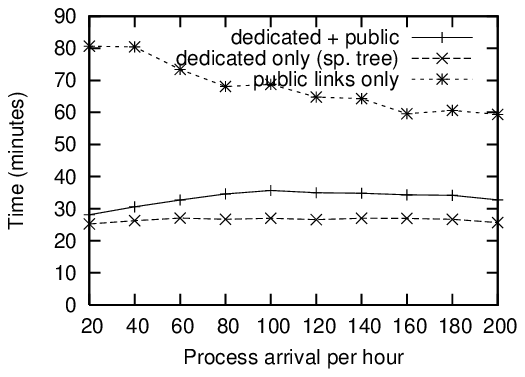}
  }
  \caption{Comparing bi-modal and uni-modal networks}
\end{figure*}

Unless otherwise mentioned, we assumed the platform to have $100$
server nodes and $99$ dedicated links interconnecting them. There were
$25$ different types of services. As the service variety is
proportional to the node degree, a node having $d$ dedicated links was
assumed to host $1+d$ different types of services (one added for
public network link). Server CPU capacity was set such that it can
execute $k$ instances of each service concurrently, according to the
mean data delivery rate. We set $k=2$. For the task workload, each
task is assumed to have $10$ service components, randomly chosen from
$25$ different types of service. Mean data delivery rate was $1$Mbps
and total amount of data to be processed from the source was $100$MB
on average. Each data point on the results shown below is an average
of $100$ observations from different experiments on randomly generated
networks with specified parameters. For each experiment, a synthetic
workload trace containing $500$ stream processing tasks were
generated. The task arrival process is assumed to be Poisson, with the
arrival rate varying across the experiments. If not mentioned
otherwise, the default arrival rate was $60$ tasks per hour.

\subsection{Benefits of Combining Opportunistic and Dedicated Resources}
We performed several sets of experiments to evaluate the benefits of
using bi-modal networks for stream processing tasks.
In  the experiments, we compare three possible settings
-- i) a network with the dedicated links only, ii) public network
only, and iii) a network that combines both. 

First argument in favor of a bi-modal network for stream processing is
that combining the public network with dedicated links, the system
achieves much higher work throughput at the same cost. To examine
this, we fed similar workload traces under same arrival rates to two
system set-ups, one with only dedicate link based networks and the
other using the combination of dedicated links and public
network. From Figure~\ref{fig:netdyna_acceptratio_numlinks} we observe
that for the same workload, if the platform uses dedicated links only,
it needs more than $120$ links to get $50\%$ acceptance ratio, whereas
the same acceptance ratio can be obtained with $50$ dedicated links
only, if the public network is utilized in conjunction. As a result,
the bi-modal system yields much higher work throughput for same number
of dedicated links.

The next argument is that utilization of the privately deployed
expensive dedicated resources such as servers and dedicated links is
increased, if inexpensive public network is used in conjunction.  From
Figure~\ref{fig:netdyna_cpuutilnet_arvrate} we observe that when a
combination of dedicated links and the public network is used, the
server utilization is much higher than the utilization when only one
type of link is used. The synergy of bi-modal links is evident here,
because, when sufficiently loaded, the server utilization of bi-modal
system is higher than the sum of utilizations in the two other cases.

Figures~\ref{fig:netdyna_dedutilnet_arvrate}
and~\ref{fig:netdyna_dedutilnet_numlinks} show another evidence of
higher return on investment. In
Figure~\ref{fig:netdyna_dedutilnet_arvrate}, we observe that the
utilization of dedicated links becomes consistently higher across a
wide range of loading scenarios if the public network is used in
combination. The lower utilization in case of a dedicated link only
network results from the fact that the platform has rejected many task
requests that would have been feasible by the augmentation of the
public resources. Figure~\ref{fig:netdyna_dedutilnet_numlinks} shows
the variation of utilization of the dedicated links with the number of
dedicated links. We observe that the difference in utilization
diminishes as the number of installed links increases. This is because
when there is sufficient number of dedicated links to carry the
required traffic of all the tasks, the public resources are not used
at all, and the bi-modal system becomes equivalent to a dedicated link
only system. In both cases, utilization of the links keeps decreasing
when more and more links are added because the workload is held
constant.

Next, we investigate how the bi-modal network helps the stream
processing platform to keep the compliance with the services contracts
it has with individual tasks. We measure the compliance as follows.
Each task request specifies a time window $T$ that is used to monitor
the delivery rate. We measured the deviation from the required rate as
$\sum_{\mbox{over all windows}}\frac{B - \hat{B}}{B}$, where $B$ is
the desired rate and $\hat{B}$ is the observed rate of delivery. In
Figure~\ref{fig:netdyna_deviation_numlinks}, we observe that the
deviation in the bi-modal system gets closer to zero as more and more
dedicated links are added to the network. However, beyond certain
number of links, ($125$ in this particular experiment), the
improvement is very marginal.  Note that deviation is counted on the
accepted jobs only. So, even though for a dedicated link only network,
the deviation is almost zero, we have seen that such network is unable
to accept enough jobs to fully utilize the resources.

When we use a combination of dedicated and public links, it is
expected that the completion time of each task will be slightly
elongated compared to a system with only dedicated links, due to the
variability in the public network. Nevertheless, using the combination
contains the elongation to a small value, compared to the case where
only public network is available. In
Figure~\ref{fig:netdyna_exectime_arvrate}, we observe a $10-20\%$
increase in the execution time in the bi-modal system, whereas
execution time would be $200-300\%$ more in case of a public network
only system.

\subsection{Necessity of Periodic Re-Scheduling}

\begin{figure*}[htb]
  \centering
  \setcounter{subfigure}{0}
    \subfigure[Throughput]{
          \label{fig:netdyna_throughput_arvrate_comparesched}
          \includegraphics[scale=0.98]{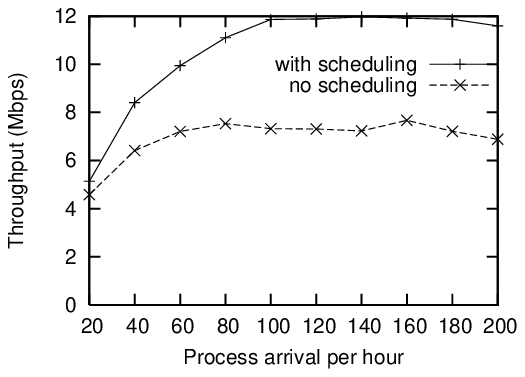}
    }
    \subfigure[Utilization of dedicated links]{
          \label{fig:netdyna_dedutilnet_arvrate_comparesched}
          \includegraphics[scale=0.98]{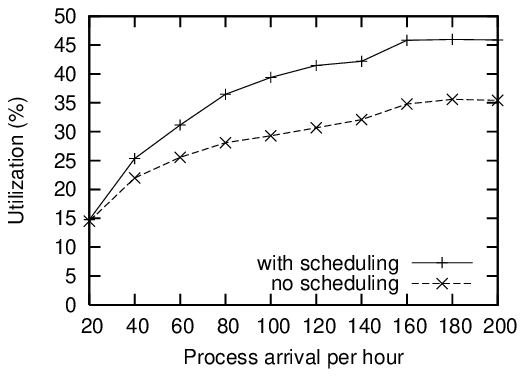}
    }
    \subfigure[Deviation from target rate]{
          \label{fig:netdyna_deviation_arvrate_comparesched}
          \includegraphics[scale=0.98]{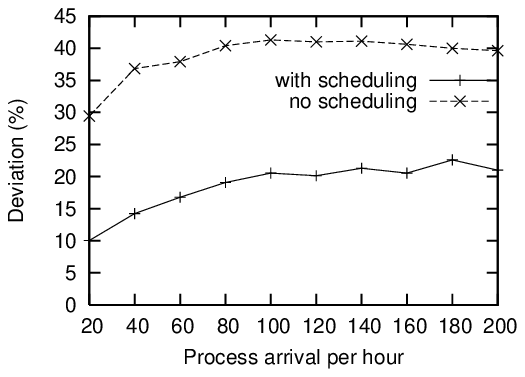}
    }
    \caption{Effect of dynamic scheduling}
\end{figure*}

Another important question in managing the bi-modal stream processing
platform is the importance of dynamic re-allocation of network links.
The main intuition behind introducing dynamic re-allocation is that
the flows that goes through the public network suffer from the
variability and lag from the target rate, whereas the flows that uses
dedicated links all-through, do not lag from the target at
all. Dynamic scheduling introduces fairness across all the tasks.  So
if link assignment is done dynamically, it is expected to improve the
utilization of the resources and increase the overall capacity of the
system.

We fed the same workload to two system set-ups containing combinations
of dedicated links and public network links. In one we disabled
dynamic re-scheduling of links and let the tasks complete with the
initial assignment of links and nodes.  From
Figures~\ref{fig:netdyna_throughput_arvrate_comparesched} we observe
that overall system throughput increases with dynamic scheduling, as
an indication of higher task acceptance ratio and higher utilization
of the system resources.
Figure~\ref{fig:netdyna_dedutilnet_arvrate_comparesched} demonstrates
that dynamic scheduling results in much higher utilization of the
dedicated links. CPU utilization remains unchanged (not shown),
because the dynamic re-allocation does not alter the node
assignments. Another rationale behind re-allocations is to increase
fairness and improve compliance with the target delivery
rate. Figure~\ref{fig:netdyna_deviation_arvrate_comparesched} shows
that irrespective of workload, the dynamic scheduling decreases the
deviation from the specified target, having the same number of
dedicated links and same public network bandwidth.

\section{Related Work}
\label{sec:related}

Architectures and resource management schemes for distributed stream
processing platforms have been studied by many research groups from
distributed databases, sensor networks, and multimedia streaming.  In
database and sensor network research, the major focus was placing the
query operators to nodes inside the network that carries the data
stream from source to the viewer~\cite{Pietzuch2006}. In multimedia
streaming problems, similar requirements arise when we need to perform
a series of on-line operations such as trans-coding or embedding on
one or more multimedia streams and these services are provided by
servers in distributed locations. In both cases, the main problem is
to allocate the node resources where certain processing need to be
performed along with the network bandwidths that will carry the data
stream through these nodes.

Finding the optimal solution to this resource allocation problem is
inherently complex. Several heuristics have been proposed in the
literature to obtain near-optimal solutions.  Recursive partitioning
of the network of computing nodes have been proposed
in~\cite{Schwan2005} and~\cite{Seshadri2007} to map the stream
processing operators on a hierarchy of node-groups.
In~\cite{Baochun2004} and~\cite{Gu2006}, the service requirements for
multi-step processing of multimedia streams have been mapped to an
overlay network of servers after pruning the whole resource network
into a subset of compatible resources. The mapping is performed
subject to some end-to-end quality constraints, but the CPU
requirements for each individual service component is not considered.
Liang and Nahrstedt in~\cite{Nahrstedt2006} have proposed solutions to
the mapping problem where both node capacity requirement and bandwidth
requirements are fulfilled. However, their proposed solution requires
global knowledge of network topolgy at a single node.

In all of the abovementioned works, the operator nodes are assumed to
interconnected through an application dependent overlay network using
the Internet as underlay. In~\cite{Nahrstedt2003}, Gu and Nahrstedt
presented a service overlay network for multimedia stream processing,
where they have shown that dynamic re-allocation of the operator nodes
provides better compliance with the service contracts in terms of
service availability and response time. However, none of the works
have proposed the use of dedicated links in conjunction with public
Internet for improving adherence to the service contracts.

\section{Conclusion}
\label{sec:conc}

In this paper, we investigated the resource management problem with
regard to data stream processing tasks. In particular, we examined how
a hybrid platform made up of dedicated server resources and bi-modal
network resources (dedicated plus public) can be used for this class
of applications. From the simulation based investigations, we were
able make several interesting observations. First, bi-modal networks
can improve dedicated resource utilization (server plus dedicated
network links). This means higher return on investment can be obtained
by engaging the bi-modal network. Second, the overall system is able
to admit and process tasks at a higher rate compared to system
configurations that do not leverage a bi-modal network. Because the
public network is engaged at zero or very low cost, this improvement
in throughput can be result in significant economic gain for
institutions that perform data stream processing workloads. Third, the
engagement of bi-modal network comes at a slight overhead that adds
small delays in stream processing tasks. Compared to public-only
networks the delays provided by the bi-modal network is almost
negligible. Fourth, dynamic rescheduling is essential to cope with
varying network conditions -- particularly in the public network. The
dynamic rescheduling algorithm switches the flows according to the
recomputed priority values to achieve the best service level
compliances.

In summary, our study highlights the benefits of the bi-modal
architecture for compute and communication-intensive
applications. Moreover, it provides simple distributed algorithms that
allows the effective utilization of such a platform for data stream
processing applications.

\singlespace

%
\balance
\bibliographystyle{latex8}
\bibliography{asad}

\begin{thebibliography}{10}\setlength{\itemsep}{-1ex}\small

\bibitem{AsadPhDThesis}
S.~Asaduzzaman.
\newblock {\em {Managing Opportunistic and Dedicated Resources in a Bi-modal
  Service Deployment Architecture}}.
\newblock PhD thesis, {School of Computer Science, McGill University}, Jan.
  2008.

\bibitem{JPDC06}
S.~Asaduzzaman and M.~Maheswaran.
\newblock {Utilizing Unreliable Public Resources for Higher Profit and Better
  SLA Compliance in Computing Utilities}.
\newblock {\em Journal of Parallel and Distributed Computing}, 66(6):796--806,
  2006.

\bibitem{jpdc07}
S.~Asaduzzaman and M.~Maheswaran.
\newblock {Strategies to Create Platforms for Differentiated Services from
  Dedicated and Opportunistic Resources}.
\newblock {\em Journal of Parallel and Distributed Computing},
  67(10):1119--1134, 2007.

\bibitem{HPCS2007}
S.~Asaduzzaman and M.~Maheswaran.
\newblock Towards a decentralized algorithm for mapping network and
  computational resources for distributed data-flow computations.
\newblock In {\em 21st IEEE HPCS}, page~30, May 2007.

\bibitem{Barabasi1999}
A.~Barabasi and R.~Albert.
\newblock {Emergence of Scaling in Random Networks}.
\newblock {\em Science}, 286(5439):509--512, 1999.

\bibitem{Jist2005}
R.~Barr, Z.~J. Haas, and R.~van Renesse.
\newblock {JiST: An efficient approach to simulation using virtual machines}.
\newblock {\em Software: Practice and Experience}, 35(6):539--576, 2005.

\bibitem{Gates2004}
L.~Chen, K.~Reddy, and G.~Agrawal.
\newblock {GATES: A Grid-Based Middleware for Processing Distributed Data
  Streams}.
\newblock In {\em HPDC}, pages 192--201, Jul. 2004.

\bibitem{Kalogeraki2007}
Y.~Drougas and V.~Kalogeraki.
\newblock {RASC: Dynamic Rate Allocation for Distributed Stream Processing
  Application}.
\newblock In {\em IPDPS}, Mar. 2007.

\bibitem{Boutaba2008}
W.~Golab and R.~Boutaba.
\newblock {Path Selection in User-controlled Circuit-switched Optical
  Networks}.
\newblock {\em Optical Switching and Networking}, 5(2-3):123--138, Jun. 2008.

\bibitem{Gu2006}
X.~Gu and K.~Nahrstedt.
\newblock Distributed multimedia service composition with statistical {QoS}
  assurances.
\newblock {\em IEEE Trans.\ Multimedia}, 8(1):141--151, 2006.

\bibitem{Nahrstedt2003}
X.~Gu, K.~Nahrstedt, R.~N. Chang, and C.~Ward.
\newblock Qos-assured service composition in managed service overlay networks.
\newblock In {\em ICDCS}, pages 194--203, May 2003.

\bibitem{Karamcheti2004}
T.~Kichkaylo and V.~Karamcheti.
\newblock {Optimal Resource-Aware Deployment Planning for Component-based
  Distributed Applications}.
\newblock In {\em HPDC}, pages 150--159, Jul. 2004.

\bibitem{Schwan2005}
V.~Kumar, B.~F. Cooper, Z.~Cai, G.~Eisenhauer, and K.~Schwan.
\newblock Resource aware distributed stream management using dynamic overlays.
\newblock In {\em Proc.\ 25th IEEE ICDCS}, pages 783--792, Jun. 2005.

\bibitem{Nahrstedt2006}
J.~Liang and K.~Nahrstedt.
\newblock Service composition for generic service graphs.
\newblock {\em Multimedia Systems}, 11(6):568--581, 2006.

\bibitem{STREAM2003}
R.~Motwani, J.~Widom, A.~Arasu, B.~Babcock, S.~Babu, M.~Datar, G.~Manku,
  C.~Olston, J.~Rosenstein, and R.~Varma.
\newblock {Query Processing, Resource Management, and Approximation in a Data
  Stream Management System}.
\newblock In {\em CIDR-2003}, Jan. 2003.

\bibitem{Pietzuch2006}
P.~R. Pietzuch, J.~Ledlie, J.~Shneidman, M.~Roussopoulos, M.~Welsh, and M.~I.
  Seltzer.
\newblock {Network-Aware Operator Placement for Stream-Processing Systems}.
\newblock In {\em ICDE}, page~49, Apr. 2006.

\bibitem{Seshadri2007}
S.~Seshadri, V.~Kumar, B.~F. Cooper, and L.~Liu.
\newblock {Optimizing Multiple Distributed Stream Queries Using Hierarchical
  Network Partitions}.
\newblock In {\em IPDPS}, pages 1--10, Mar. 2007.

\bibitem{Wallerich2006}
J.~Wallerich and A.~Feldmann.
\newblock Capturing the variability of internet flows across time.
\newblock In {\em INFOCOM}, Apr. 2006.

\bibitem{Baochun2004}
M.~Wang, B.~Li, and Z.~Li.
\newblock {sFlow: Towards resource-efficient and agile service federation in
  service overlay networks}.
\newblock In {\em Proc.\ 24th IEEE ICDCS}, pages 628--635, Mar. 2004.

\end{thebibliography}

\end{document}